\documentclass[12pt, onecolumn, notitlepage]{article}
\usepackage{graphicx}
\usepackage{amsmath}
\usepackage{amssymb}
\usepackage{authblk}
\usepackage{float}

\usepackage{fixltx2e}

\allowdisplaybreaks

\begin{document}

\title{Stationary States in Bistable System Driven 
by L\'evy Noise}

\author[1]{O.Yu. Sliusarenko\thanks{aslusarenko@kipt.kharkov.ua}}
\author[2]{D.A. Surkov}
\author[1]{V.Yu. Gonchar}
\author[1,3]{A.V. Chechkin}
\affil[1]{Akhiezer Institute for Theoretical Physics National Science Center ``Kharkiv Institute of Physics and Technology'', Akademicheskaya st.~1, Kharkiv 61108, Ukraine}
\affil[2]{Karazin National University, Svobody sq.~4, Kharkiv 61000, Ukraine}
\affil[3]{Institute for Physics and Astronomy University of Potsdam, 14476 Potsdam-Golm, Germany}
\abstract{
We study the properties of the probability density
function (PDF) of a bistable system driven by heavy tailed 
white symmetric L\'evy noise. The shape of the
stationary PDF is found analytically for the particular case
of the L\'evy index $\alpha = 1$ (Cauchy noise). For an
arbitrary L\'evy index we employ numerical methods based
on the solution of the stochastic Langevin equation and 
space fractional kinetic equation. In contrast with the 
bistable system driven by Gaussian noise, 
in the L\'evy case the positions of maxima of the stationary PDF 
do not coincide with the positions of minima of the 
bistable potential. We provide a detailed study of the
distance between the maxima and the minima as a function
of the potential's depth and L\'evy noise parameters.
} 
\maketitle
\section{Introduction}
\label{sec.intro}

The term ``L\'evy noise'' stands for a class of non-Gaussian random stationary processes possessing
alpha-stable L\'evy probability distributions, the class of probability laws originally
studied by French mathematician Paul Pierre L\'evy \cite {levy1954theorie}. These laws are
ubiquitous in nature due to generalized central limit theorem \cite{gnedenko1968limit}.
The alpha-stable probability density functions (PDFs) exhibit slowly decaying power-law
asymptotic behavior of the form $|x|^{-1-\alpha}$, where $\alpha$ is called the L\'evy
index, $0 < \alpha < 2$; therefore, the variance is infinite. For that reason
L\'evy noises naturally arise in the description of random processes with large outliers,
far from equilibrium, see, e.g., \cite{janicki1994simulation,nikias1995signal,klages2008anomalous,meyers2009encyclopedia} and references therein.  

The study of relaxation processes in dynamical systems subjected to the L\'evy noise
is of interest for several reasons. 
From the nonequilibrium statistical physics point of view such systems represent 
paradigmatic examples of non-Brownian
processes in the systems far from equilibrium. Also, it has been demonstrated that
such systems can be used as ``minimal'' models for the description of such diverse
phenomena as anomalous transport in turbulent plasmas \cite{chechkin2002fractional,jha2003evidence,gonchar2003stable,mizuuchi2005edge} and abrupt
climatic changes \cite{ditlevsen1996contrasting,ditlevsen1999anomalous,ditlevsen1999grl}. Another important field of research is related
to the reverse engineering problem \cite{eliazar2003levy}.

The properties of linear dynamical systems subjected to the L\'evy noise have been studied
in the two seminal papers \cite{west1982linear,peseckis1987statistical} and later in \cite{jespersen1999levy,CheGo}.
The relaxation in nonlinear systems has been extensively studied within the last
decade, see the reviews \cite{klages2008anomalous,meyers2009encyclopedia,dubkov2008levy} and references therein, and more
recently, e.g., in \cite{dybiec,denisov2008steady,dybiec2010stationary,pavlyukevich2010levy,dubkov2009problem,la2010dynamics}.  

In the present short communication we demonstrate the first results on stationary states
in nonlinear dynamical system with a symmetric double-well potential. We find
analytical expression for the stationary PDF for the Cauchy case $\alpha = 1$ and employ 
numerical methods to investigate properties of the PDFs for arbitrary $\alpha$.

\section{Analytical approach}
\label{sec.analytics}

Our starting point is the stochastic Langevin equation which can be written
in dimensionless variables as 
\begin{equation} 
\label{eq.init} 
\frac{dx}{dt}=-\frac{dU}{dx} + \sigma \xi_\alpha(t),
\end{equation} 
where $U(x)=x^4/4-ax^2/2$, $\xi_\alpha(t)$ is an $\alpha$-stable noise source with the unit intensity, $\sigma$ is the noise amplitude, $a>0$ is the parameter that allows to govern the potential's well depth (note, that $\pm \sqrt{a}$ are the positions of its minima). With such a choice of dimensionless variables we can compare our results with
those obtained for a quartic potential \cite {tanatarov}, setting $a=0$. This kind of a potential, along with the Langevin approach framework is used to describe a large variety of systems where a switching between two metastable states occurs, see, e.g. Refs.~\cite{ditlevsen1996contrasting,ditlevsen1999anomalous,ditlevsen1999grl} mentioned above. The corresponding to Eq.~(\ref{eq.init}) Fokker-Planck equation reads as
\cite{jespersen1999levy,CheGo}
\begin{equation}
\label {f1}
\frac {\partial f}{\partial t}=\frac {\partial}{\partial x}\left (\frac {\partial U}{\partial x}f\right )+\sigma ^{\alpha}\frac {\partial ^{\alpha}f}{\partial |x|^{\alpha}},
\end{equation}
where ${\partial ^{\alpha}f}/{\partial |x|^{\alpha}}$ is the Riesz fractional derivative \cite{samko}.

We consider the stationary case. For the characteristic function defined as $\hat  f(k,t)=\int _{-\infty}^{\infty} dxf(x,t)e^{ikx}$ we get:
\begin{equation}
\frac {d^3\hat f}{dk^3}+a\frac {d\hat f}{dk}-\sigma^\alpha |k|^{\alpha-1}\hat f =0.
\end{equation}
If we restrict ourselves to the case $\alpha =1$, the latter equation simplifies to the following:
\begin{equation}
\frac {d^3\hat f}{dk^3}+a\frac {d\hat f}{dk}-\hat f\sigma =0.
\end{equation}

Looking for the solution in the form $\hat f(k)=Ce^{zk}$, we arrive at the characteristic equation $z^3+az-\sigma=0$, for which the three roots are determined by the Cardano formulas:
$z_1=-({p+q})/{2}+i\sqrt 3({p-q})/{2}, \; z_2=z_1^*, \; z_3=p+q$, where $p=\left( { {\sigma}/{2}+\sqrt {{\sigma ^2}/{4}+ {a^3}/{27}}} \right) ^{1/3} ,\; q=\left({{\sigma}/{2}-\sqrt { {\sigma ^2}/{4}+  {a^3}/{27}}} \right)^{1/3}$, and the general solution is written as $\hat f(k)=C_1e^{zk}+C_2e^{z^*k}+C_3e^{z_3k}, \; z\equiv z_1$.

The coefficient $C_3$ is zero due to $\hat f(k)\to 0, \;k\to \infty$. The coefficients $C_1$ and $C_2$ are determined from the normalization condition $\hat f(k=0)=1$ and the boundary condition $ {df(0)}/{dk}=0$ arising due to symmetry of the problem. Thus, for the stationary state we have

\begin{equation}
\hat f(k)=\frac {1}{z-z^*}\left(ze^{z^*|k|}-z^*e^{z|k|}\right).
\end{equation}
Making an inverse Fourier transform we get the stationary PDF in case $\alpha=1$ :
\begin{equation} 
\label{eq.an}
f(x)=-\frac {|z|^2}{\pi}\frac {z+z^*}{x^4+(z^2+z^{*2})x^2+|z|^4},
\end {equation}
The minima of the $U(x)$ are located at $\pm x_U$, where $x_U = \sqrt a$. The maxima of the stationary PDF are located at $\pm x_f$, where $x_f = \sqrt {-( {z^2+z^{*2}})/{2}}$ (these maxima always exist due to $z^2+z^{*2} = -\Big[\left(\sqrt{4 a^3/27 + \sigma ^2} - \sigma\right)^{2/3}+
\left(\sqrt{4 a^3/27 + \sigma ^2} + \sigma\right)^{2/3}+
2^{8/3}a/3\Big]/2^{2/3}<0$). Clearly, $x_f$ and $x_U$ are not the same. This fact was expected from the previous studies \cite{tanatarov}, where the
   bimodality of the stationary PDF for quartic potential was revealed.

Let us introduce a parameter $\Delta=(x_f-x_U)/{x_U}$. In the weak noise limit, $ {a^3}/{\sigma ^2} \gg 1$, we get $\Delta \propto {\sigma ^2}/{a^3}$. In the strong noise limit, ${a^3}/{\sigma ^2} \ll 1 $, one may become convinced that $x_f\approx  {\sigma ^{{1}/{3}}}/{\sqrt 2}$ and  $\Delta \approx \left ({\sigma ^2}/{a^3}\right)^{{1}/{6}}/{\sqrt2}-1$. The last expression reveals that the distance between the corresponding minima and maxima becomes unlimited with the increasing noise intensity.

\section{Numerical results}
\label{sec.numerics}

In this section we employ two different techniques to simulate the 
stationary PDFs. The first method is based on the numerical solution of
space-fractional Fokker-Planck equation by means of the Gr\"{u}nwald-Letnikov 
approximation of the Riesz fractional derivative \cite {chechkin2004levy}. 
The second method uses direct Langevin dynamics simulations.

Since both these approaches give us stationary PDF it is straightforward to compare their results to each other, as well as to the analytical results obtained in Section~\ref{sec.analytics}. In the kinetic approach we investigate the system by varying potential's parameter $a$, while focusing on the noise intensity properties in the Langevin simulations instead.

\begin{figure*}[h!]
\centering
\includegraphics[width=0.99\linewidth]{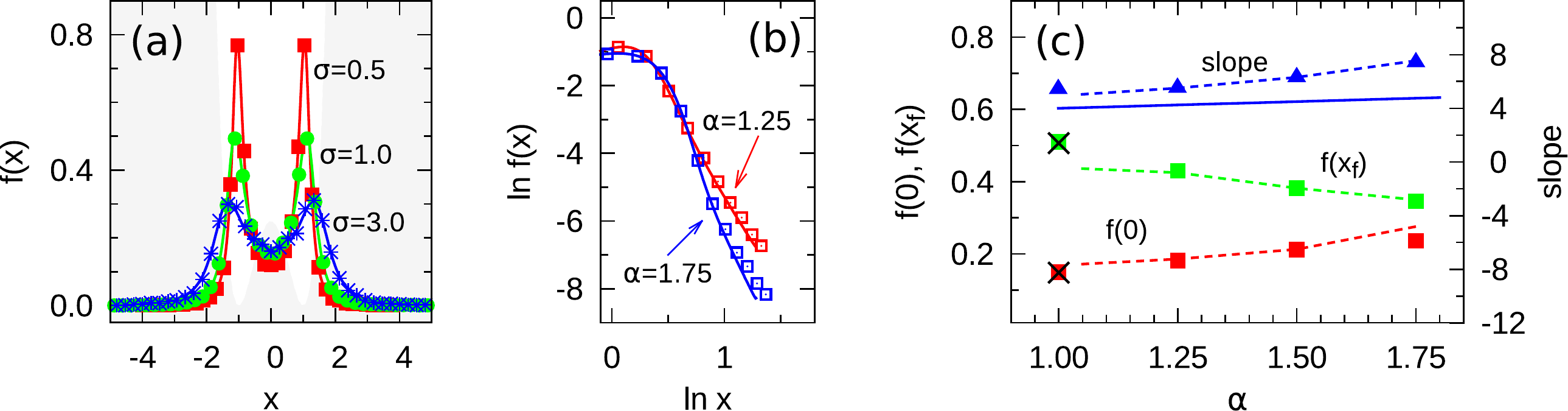}
\caption{\label{fig.pdfLevy}\textbf{(colour online)} (a): Comparison of analytical (solid line, Eq.~\ref{eq.an}) and numerical (dots) results for stationary PDF in double-well potential, $a=1$, $\alpha=1$. (b): Comparison of Langevin and kinetic 
simulations results. Shapes of PDFs in linear and log-log scales, $a=\sigma = 1$. (c): Quantitative comparison of the shape parameters, $a=\sigma = 1$.}
\end{figure*}

As the first test of the approaches, let us compare the analytical results obtained in Section~\ref{sec.analytics} with the corresponding numerical data of the Langevin simulations as it is shown in Fig.~\ref{fig.pdfLevy}a.  Evidently, the simulations (dots) do match well with the analytics (solid lines) at all noise intensities. The Figs.~\ref{fig.pdfLevy}b and~\ref{fig.pdfLevy}c show the results of both simulation methods, the Langevin (points) and the kinetic (solid or dashed lines) for different values of the L\'evy index. 

The double logarithmic scale in Fig.~\ref{fig.pdfLevy}b makes it obvious that the slopes of the asymptotes have a disrepancy which is probably due to a growing inaccuracy in extracting PDF from of the Langevin simulations at long distances. Fig.~\ref{fig.pdfLevy}c gives a quantitative comparison of the methods' results plotting the stationary PDF peaks $f(x_f)$ and minima $f(0)$, and the asymptotes slopes $slope=(-d \ln f/d\ln x), \, x \to \infty$ showing a good agreement between the results obtained with the use of both simulation techniques. In Fig.~\ref{fig.pdfLevy}c we also plot two analytical results. The values of $f(x_f)$ and $f(0)$ computed from Eq.~(\ref{eq.an}) are shown by two black crosses. The blue solid line shows the asymptotic slope of the stationary PDF $f(x) \propto 1/|x|^{\alpha + 3}$ (see Eq.~(71) in \cite{chechkin2004levy}) obtained for a quartic potential well. The discrepancy between this slope and those obtained with numerical simulations (dashed line and triangles) indicates the necessity of taking into account the quadratic term while estimating the asymptotic properties of the stationary PDF in the double-well potential. 

\begin{figure*}[tbp]
\centering
\includegraphics[width=0.8\linewidth]{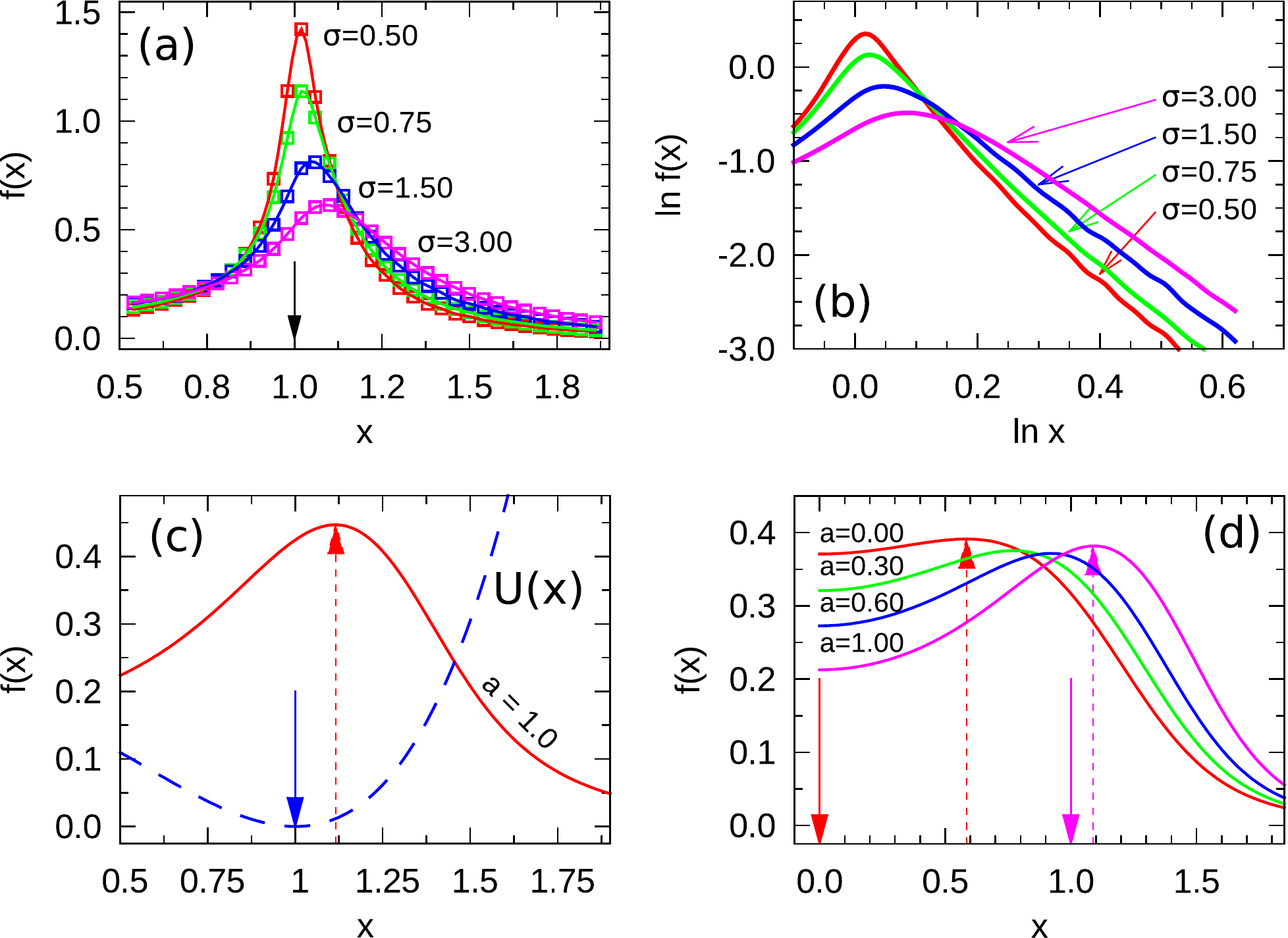}
\caption{\label{fig.pdfLevy3}\textbf{(colour online)} (a,b): Langevin simulations of stationary PDFs in linear and log-log scales, respectively. Case $\alpha=0.5, \, a=1$. (c,d): Kinetic simulations of stationary PDFs. (c): $\alpha = 1.05$, $a=1.0$, $\sigma=1$; (d): $\alpha=1.50$, $\sigma=1$. Arrows show the positions of the potential minimum's location $\sqrt{a}$ ($\downarrow$) and the PDF's peak coordinate $x_f$ ($\uparrow$).}
\end{figure*}

\begin{figure*}[tbp]
\centering
\includegraphics[width=0.8\linewidth]{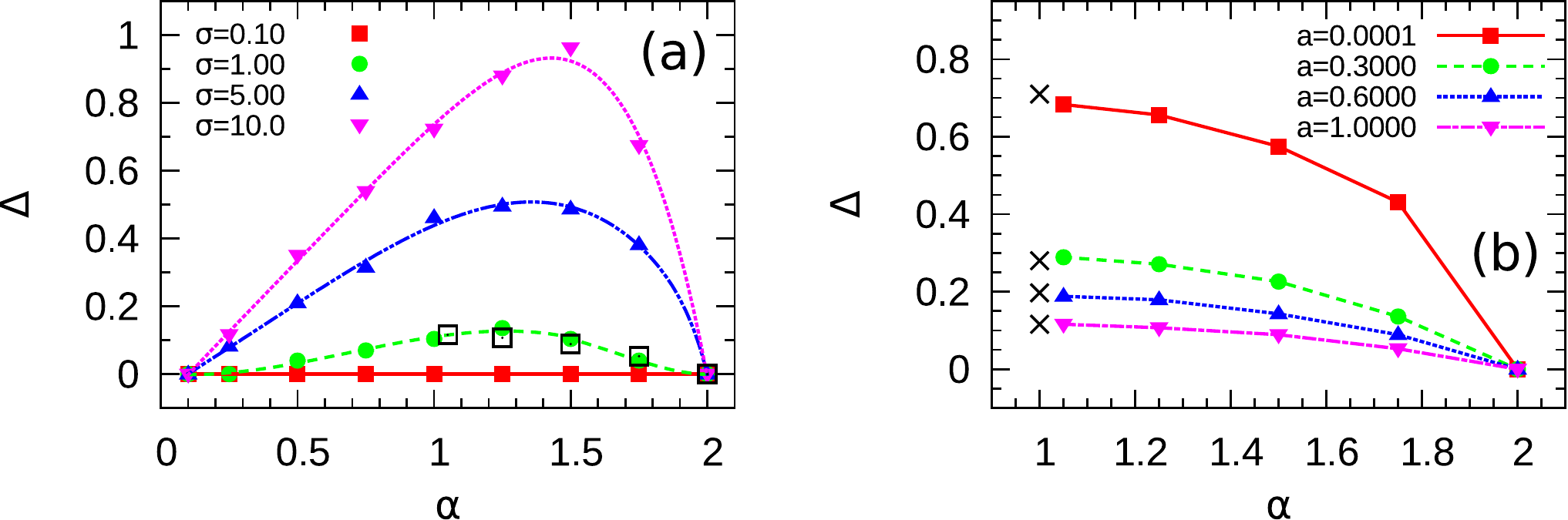}
\caption{\label{fig.pdfLevy5}\textbf{(colour online)} Quantitative analysis of the
distance between the PDF peaks and potential minima. (a): The results of Langevin simulations for various noise intensities $\sigma$ and $a=1$ are shown with filled squares and triangles; with empty $\square$ we depict the data obtained via the numerical kinetic approach for $\sigma=1$. (b): The results of the kinetic approach for various coefficients $a$ alongside with analytical results marked with $\times$.}
\end{figure*}

In Fig.~\ref{fig.pdfLevy3} the shapes of stationary PDFs are shown for different values of the parameters $\alpha$, $a$ and $\sigma$. The main feature of the stationary PDFs is demonstrated here clearly: the PDF's peaks do not coincide with potential's minima.  
The results of quantitative analysis of this phenomenon is presented in Fig.~\ref{fig.pdfLevy5} in terms of the coefficient $\Delta$. As it should be, $\Delta \to 0$ at $\alpha \to 2$, however, $\Delta \neq 0$ at other L\'evy indices of the noise. The deviation becomes more visible at higher noise intensities and when the potential's minimum becomes more shallow. Interestingly, our numerical simulations 
demonstrate the existence of maximum at intermediate $\alpha$'s in Fig.~\ref{fig.pdfLevy5}a. Both figures show a good agreement of the results obtained numerically and analytically.

\section{Conclusions}

Our studies of the L\'evy flights in a double-well potential demonstrate that the 
positions of the maxima of the stationary probability density functions do not coincide with the 
positions of the minima of the potential well. In the present short communication
we provide a first study of the properties of the stationary PDFs and reveal that 
the difference between the mentioned maxima and minima grows with the increasing 
noise intensity and decreases with the potential's depth. Very recently, the discrepancy 
issue between the shape of the potential
and the shape of stationary PDF was addressed from the point of view
of ``mean reversion'' and ``mode reversion'' \cite{IddoCohen}. Interestingly,
the same discrepancy was found there under certain conditions even in the 
Gaussian case. We plan to elaborate on the ``reversion'' phenomena
for L\'evy flights in the forthcoming long paper.



\begin{thebibliography}{10}
\bibitem{levy1954theorie}
P.~L{\'e}vy,
\newblock {\em Th{\'e}orie de l'addition des variables al{\'e}atoires},
  volume~1
\newblock (Gauthier-Villars, Paris:, 1954).
\bibitem{gnedenko1968limit}
B.~Gnedenko, A.~Kolmogorov, K.~Chung, and J.~Doob,
\newblock {\em Limit distributions for sums of independent random variables},
  volume 195
\newblock (Addison-Wesley Reading, MA:, 1968).
\bibitem{janicki1994simulation}
A.~Janicki and A.~Weron,
\newblock {\em Simulation and chaotic behavior of $\alpha$-stable stochastic
  processes}, volume 178
\newblock (CRC, 1994).
\bibitem{nikias1995signal}
C.~Nikias and M.~Shao,
\newblock {\em Signal processing with alpha-stable distributions and
  applications}
\newblock (Wiley-Interscience, 1995).
\bibitem{klages2008anomalous}
A. Chechkin, R. Metzler, J. Klafter, V. Gonchar, Introduction to
the Theory of L\'evy Flights. In R. Klages, G. Radons, I.M. Sokolov (Eds),
\newblock {\em Anomalous Transport: Foundations and Applications}
\newblock (Wiley-VCH, 2008, PP. 129 - 162.)
\bibitem{meyers2009encyclopedia}
R.~Metzler, A.V.~Chechkin, and J.~Klafter, In R.A. Mayers (Ed), 
\textit{Encyclopedia of Complexity and System Science}, Article 293
\newblock (Springer-Verlag, Berlin, 2009).
\bibitem{chechkin2002fractional}
A.~Chechkin, V.~Gonchar, and M.~Szyd{\l}owski,
\newblock Physics of Plasmas {\bf 9}, 78 (2002).
\bibitem{jha2003evidence}
R.~Jha, P.~Kaw, D.~Kulkarni, J.~Parikh, and A.~Team,
\newblock Physics of Plasmas {\bf 10}, 699 (2003).
\bibitem{gonchar2003stable}
V.~Gonchar et~al.,
\newblock Plasma Physics Reports {\bf 29}, 380 (2003).
\bibitem{mizuuchi2005edge}
T.~Mizuuchi et~al.,
\newblock Journal of Nuclear Materials {\bf 337}, 332 (2005).
\bibitem{ditlevsen1996contrasting}
P.~Ditlevsen, H.~Svensmark, and S.~Johnsen,
\newblock Nature {\bf 379}, 810 (1996).
\bibitem{ditlevsen1999anomalous}
P.~Ditlevsen,
\newblock Physical Review E {\bf 60}, 172 (1999).
\bibitem{ditlevsen1999grl}
P.D.~Ditlevsen, 
Geophys. Res. Lett. {\bf 26}, 1441 (1999).
\bibitem{eliazar2003levy}
I.~Eliazar and J.~Klafter,
\newblock Journal of Statistical Physics {\bf 111}, 739 (2003).
\bibitem{west1982linear}
B.~West and V.~Seshadri,
\newblock Physica A: Statistical and Theoretical Physics {\bf 113}, 203 (1982).
\bibitem{peseckis1987statistical}
F.~Peseckis,
\newblock Physical Review A {\bf 36}, 892 (1987).
\bibitem{jespersen1999levy}
S.~Jespersen, R.~Metzler, and H.~Fogedby,
\newblock Physical Review E {\bf 59}, 2736 (1999).
\bibitem{CheGo}
A.~Chechkin and V.~Gonchar,
\newblock J. Eksper. Theor. Phys. {\bf 91}, 635 (2000).
\bibitem{dubkov2008levy}
A.~Dubkov, B.~Spagnolo, and V.~Uchaikin,
\newblock Int. J. Bifur. Chaos {\bf 18}, 2649 (2008).
\bibitem{dybiec}
B.~Dybiec, E.~Gudowska-Nowak, and I.M.~Sokolov, 
\newblock Physical Review E {\bf 76}, 041122 (2007).
\bibitem{denisov2008steady}
S.~Denisov, W.~Horsthemke, and P.~H{\"a}nggi,
\newblock Physical Review E {\bf 77}, 061112 (2008).
\bibitem{dybiec2010stationary}
B.~Dybiec, I.~Sokolov, and A.~Chechkin,
\newblock Journal of Statistical Mechanics: Theory and Experiment {\bf 2010},
  P07008 (2010).
\bibitem{pavlyukevich2010levy}
I.~Pavlyukevich, B.~Dybiec, A.~Chechkin, and I.~Sokolov,
\newblock The European Physical Journal-Special Topics {\bf 191}, 223 (2010).
\bibitem{dubkov2009problem}
A.~Dubkov, A.~La~Cognata, and B.~Spagnolo,
\newblock Journal of Statistical Mechanics: Theory and Experiment {\bf 2009},
  P01002 (2009).
\bibitem{la2010dynamics}
A.~La~Cognata, D.~Valenti, A.~Dubkov, and B.~Spagnolo,
\newblock Physical Review E {\bf 82}, 011121 (2010).
\bibitem{tanatarov}
A.~Chechkin, V.~Gonchar, J.~Klafter, R.~Metzler, and L.~Tanatarov,
\newblock Chem. Phys. {\bf 284}, 233 (2002).
\bibitem{samko}
S.G.~Samko, A.A. Kilbas, and O.I. Marichev, 
\newblock {\em Fractional Integrals and Derivatives: Theory and applications}
\newblock (Gordon and Breach, PA:, 1993).
\bibitem{chechkin2004levy}
A.~Chechkin, V.~Gonchar, J.~Klafter, R.~Metzler, and L.~Tanatarov,
\newblock Journal of Statistical Physics {\bf 115}, 1505 (2004).
\bibitem{IddoCohen}
I.I.~Eliazar and M.H.~Cohen, J. Phys. A: Math. Theor. {\bf 45}, 332001 (2012).


\end{thebibliography}
\end{document}